\documentclass[notitlepage,amsmath,amssymb,aps,prl,twocolumn,nofootinbib,superscriptaddress]{revtex4-1}
\usepackage{physics}
\usepackage{slashed}
\usepackage{graphicx}
\usepackage{xcolor}
\usepackage{xspace}
\usepackage[pdftex,final]{hyperref}
\pdfoutput=1

\newcommand{\be}{\begin{equation}}
\newcommand{\ee}{\end{equation}}

\newcommand{\Nf}{N_\mathrm{f}}
\newcommand{\psib}{\bar{\psi}}

\begin{document}

\author{Charlie Cresswell-Hogg}
\email{charles.cresswell@tu-dortmund.de}
\email{c.cresswell-hogg@sussex.ac.uk}
\affiliation{Department of Physics and Astronomy, University of Sussex, Brighton, BN1 9QH, U.K.}
\affiliation{Department of Physics, TU Dortmund, Otto-Hahn-Straße 4, 44227 Dortmund, Germany}

\author{Daniel F.~Litim}
\email{d.litim@sussex.ac.uk}
\affiliation{Department of Physics and Astronomy, University of Sussex, Brighton, BN1 9QH, U.K.}

\title{Four Fermi Theory in Four Dimensions is Renormalisable}

\begin{abstract}
We demonstrate the renormalisability of quantum field theories in four dimensions with elementary self-interacting Dirac fermions and to leading order in the limit of many fermion flavours $\Nf$. Starting from the underlying divergence structure and using Gross-Neveu-type interactions as a template, we explain why extended four-fermion theories including higher-derivative interactions are well-defined, renormalisable, and predictive with only a few free parameters. We also provide the exact large-$\Nf$ leading beta functions of couplings and discuss quantum scaling dimensions, universality, $1/\Nf$ corrections, and extensions to other types of four fermion interactions. Implications for effective theory and model building are indicated.
\end{abstract}

\maketitle

{\it Introduction.---} 
Perturbative renormalisability is a cornerstone of the standard model of particle physics. Together with symmetry principles, 
it  constrains the set of fundamental interactions to  those whose  {classical} scale dimensions  are lower or equal to the number of spacetime dimension $d$. Beyond the realm of perturbation theory, Wilson's notion of renormalisability additionally encompasses  interactions whose short-distance {\it quantum} scale dimensions   are lower or equal to $d$  \cite{Wilson:1971dh,Wilson:1973jj}.
If so, canonically non-renormalisable interactions  may remain   well-defined  due to residual interactions at high energies, opening up  new avenues for model building.

The poster child for Wilson's  scenario are four-fermion (4F) interactions in three dimensions. Even though non-renorma\-lisable  by  canonical power counting, 4F theories have been shown to be renormalisable, predictive, and ultraviolet (UV) complete \cite{Gawedzki:1985ed,Rosenstein:1988pt,deCalan:1991km,Hands:1992be,Jakovac:2014lqa,Cresswell-Hogg:2022lgg,Cresswell-Hogg:2022lez}. Crucial for this is the existence of an interacting UV fixed point that lowers the quantum scale dimension of vertices, much in the spirit of Weinberg's asymptotic safety conjecture \cite{Weinberg:1980gg}. 

In particle physics, 4F interactions play a prominent role in model building and effective theories,~e.g.~\cite{Grzadkowski:2010es}. It is widely expected that their UV-completions require new physics starting around their respective Fermi scales. Still, renormalisation group and lattice studies have suggested that strongly coupled 4F theories in four dimensions  may   also  develop  interacting  UV fixed points \cite{Aoki:1996fh,Aoki:1999dv, Kubota:1999jf, Gies:2003dp,Gies:2005as,Hasenfratz:2022qan,Witzel:2024bly}. Unlike in three dimensions, however, strict proofs and definite conclusions are lacking \cite{Parisi:1975im,Rosenstein:1990nm,ZinnJustin:1991yn,Hands:1991py,Braun:2010tt}. 

In this Letter, we close this gap and establish the renormalisability of four-fermion theories in four dimensions starting from first principles. Using Gross-Neveu theory as a template, we identify the underlying divergences and  the necessary 
building blocks  to remedy them. Throughout, we stick to the fermionic degrees of freedom  without a need for bosonisation or  auxilliary  fields \cite{Jakovac:2014lqa,Cresswell-Hogg:2022lgg,Cresswell-Hogg:2025wda}. Strict control including at strong coupling is achieved with the help   of large-$\Nf$ resummations, with $\Nf$ the number of fermion flavours. We also study scaling dimensions of interactions, and discuss renormalisability from the viewpoint of the renormalisation group. Implications for effective theory and model building are indicated.

{\it Renormalisable  four-fermion theories.---} 
We begin  examining UV divergences in 4F theories with $\Nf$ flavours of interacting Dirac fermion $\psi_a$, focussing on Gross-Neveu type  theories \cite{Gross:1974jv} in $d$ euclidean dimensions
\be
\label{eq:localGN}
L_{\rm GN}=\psib_a \slashed{\partial} \psi_a - \tfrac{1}{2 \Nf} G ( \psib_a \psi_a )^2 \,.
\ee
The 4F coupling  is normalised  to facilitate a $1/\Nf$ expansion in standard fashion~\cite{Coleman:1980nk,Weinberg:1997rv}. Its  canonical mass dimension  $[G] = 2 - d$ is negative above two dimensions. A discrete chiral symmetry $\psi_a \mapsto \gamma^5 \psi_a$ and $\psib_a \mapsto  -\psi_a \gamma^5$ protects the generation of  mass and chirally-odd interactions. In two dimensions, the theory  is perturbatively renormalisable and asymptotically free \cite{Gross:1974jv}.

\begin{figure}
\includegraphics[width=.6\linewidth]{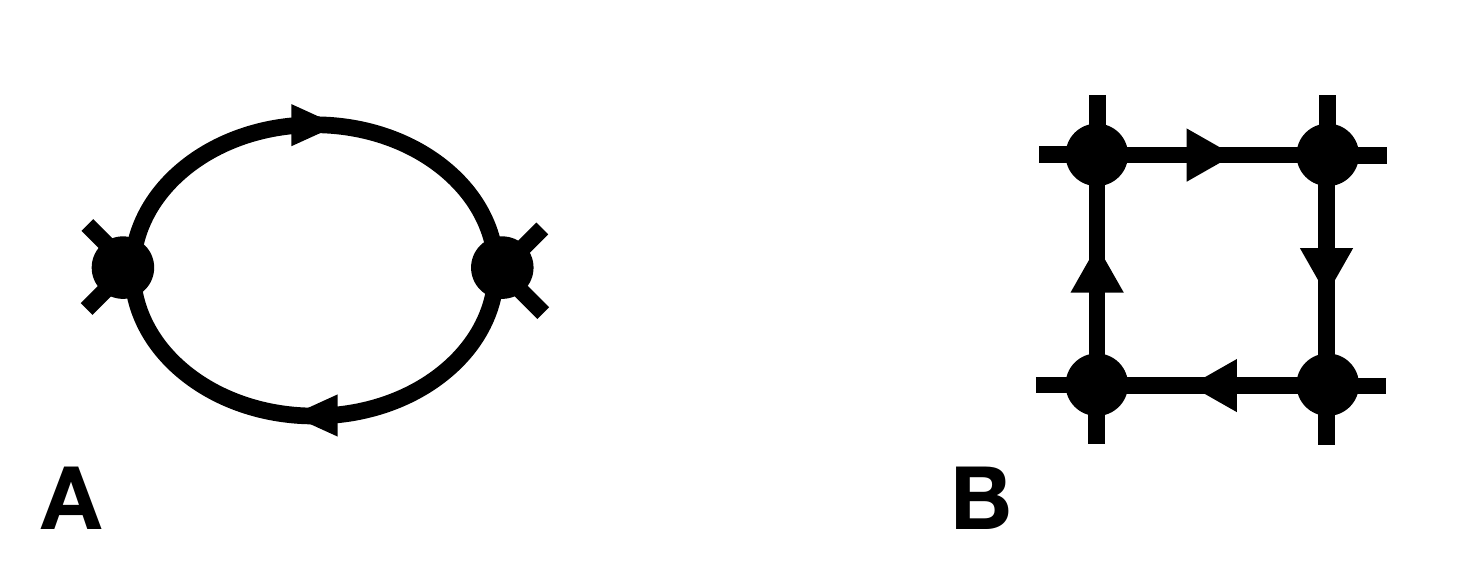}
\caption{Divergent subdiagrams appearing at leading order in the large-$\Nf$ expansion. Only the index contraction with a trace over the propagators in the loop is implied in each.}
\label{fig:subdivs}
\end{figure}

In three dimensions, the theory \eqref{eq:localGN} is renormalisable in Wilson's sense  \cite{Gawedzki:1985ed,Rosenstein:1988pt,deCalan:1991km,Hands:1992be,Jakovac:2014lqa,Cresswell-Hogg:2022lgg} and it is useful to revisit the  reason for it. Consider the large-$\Nf$ leading contributions to the fermion four-point function, given by bubble diagrams  
\be\label{eq:bubbles}
\vcenter{\hbox{\includegraphics[width=.9\linewidth]{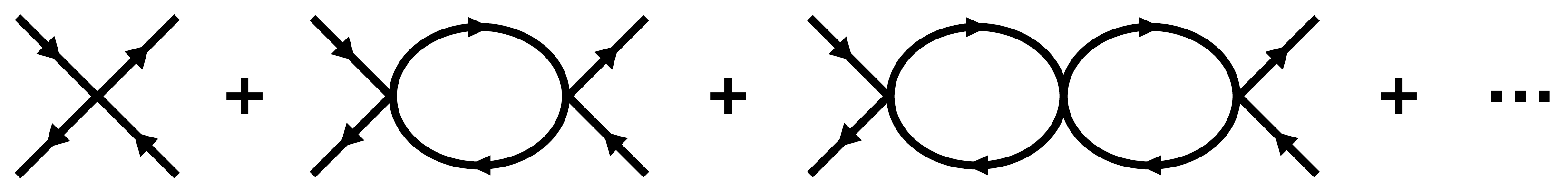}}}
\ee
that form a geometric series with repeated insertions of the one-loop bubble (Fig.~\ref{fig:subdivs}, diagram A). 
Here, only the large-$\Nf$ leading index contraction with a trace over both fermion propagators is implied. 
Summing over all terms in \eqref{eq:bubbles}, 
one obtains resummed vertices of the form 
\be\label{eq:bubbleSum}
\Nf^{-1} G \sum_{n=0}^\infty [-G B ( p^2 ) ]^n = \frac{\Nf^{-1}}{G^{-1} + B ( p^2 )} \, ,
\ee 
up to index structures and crossed terms. Each bubble contributes a factor $- \Nf B(p^2)$, where $p$ is the external momentum flowing through the loop, and each vertex gives a factor $G / \Nf$. All diagrams in the sum contribute at the same order in $1/\Nf$ if $G$ is of order unity. Regularising with a hard momentum cutoff $\Lambda$, the bubble integral is linearly divergent $B ( p^2 )\rvert_{3d} = a \Lambda+\cdots $. The theory is renormalised by absorbing the divergence into a redefinition of $G^{-1}$. Under the renormalisation group, this translates into the existence of an  interacting UV fixed point $g=G(\mu)\mu\to g_*$ that renders the 4F coupling  relevant and the theory  UV-complete \cite{Gawedzki:1985ed,Rosenstein:1988pt,deCalan:1991km,Hands:1992be,Jakovac:2014lqa,Cresswell-Hogg:2022lgg}.

{\it Four-fermion theories in four dimensions.---} 
In four dimensions, new operators beyond those given in  \eqref{eq:localGN}  become marginal in the interacting theory and  must be accounted for in the fundamental action. Diagrammatically, this  manifests itself in two new types of divergences at loop level. 
The first one relates to the bubble integral (Fig.~\ref{fig:subdivs}, diagram A), whose UV divergence now reads
\be\label{eq:Bdiv4d}
B ( p^2 )\rvert_{4d} = a \Lambda^2 + b \, p^2 \ln \Lambda + \dots
\ee
for constants $a$ and $b$. Besides  a momentum independent quadratic divergence, we observe  a  new momentum dependent logarithmic divergence $\propto p^2$. The former can be absorbed into a redefinition of the point-like 4F coupling. Via the geometric series in \eqref{eq:bubbles},  the latter implies  an infinite tower of higher-derivative 4F operators.

A second new divergence relates to the eight-fermion (8F) vertex. It originates from the  fermion box integrals (Fig.~\ref{fig:subdivs}, diagram B) that are logarithmically divergent, 
\be\label{eq:Ddiv4d}
D ( p_1, p_2, p_3 ) \rvert_{4d} = c \, \ln \Lambda + \dots
\ee
and appear to leading order as subdiagrams dressed with bubble chains. We note that $b$ and $c$ in \eqref{eq:Bdiv4d} and \eqref{eq:Ddiv4d} are universal and scheme-independent numbers, whereas $a$ is genuinely non-universal. Divergences from tadpole and triangle diagrams vanish by chiral symmetry, and counterterms for operators with more than eight fermion fields are not required. We thus conclude that renormalisability in four dimensions  necessitates  elementary 4F and 8F vertices, and  higher-derivative interactions. 

{\it Renormalising the 4F vertex.---} 
For the 4F vertex to be able to absorb the divergences  \eqref{eq:Bdiv4d}, we need to uplift the pointlike  vertex $G (\psib_a\psi_a)^2$  in  \eqref{eq:localGN}  to a quasi-local vertex
$(\psib_a\psi_a) F ( -\partial^2 ) (\psib_b\psi_b)$, where $F$ is a suitable differential operator.
In momentum space, it reads
\be\label{eq:F4d}
F ( p^2 ) = \frac{1}{Z p^2 + G^{-1}} 
\ee
and depends on the pointlike 4F coupling $G$ and a new dimensionless parameter $Z$ accounting for an infinite tower of higher-derivative 4F Mandelstam descendants of the form $\sim (\psib\psi) \partial^{2n} (\psib\psi)$.

Next, we explain how the uplift from $G$ to $F$ feeds into the resummed vertex and enables the subtraction of the divergences  \eqref{eq:Bdiv4d}. At tree-level, the vertex is given by
\be\label{eq:4Frule}\nonumber
\vcenter{\hbox{\includegraphics[height=4em]{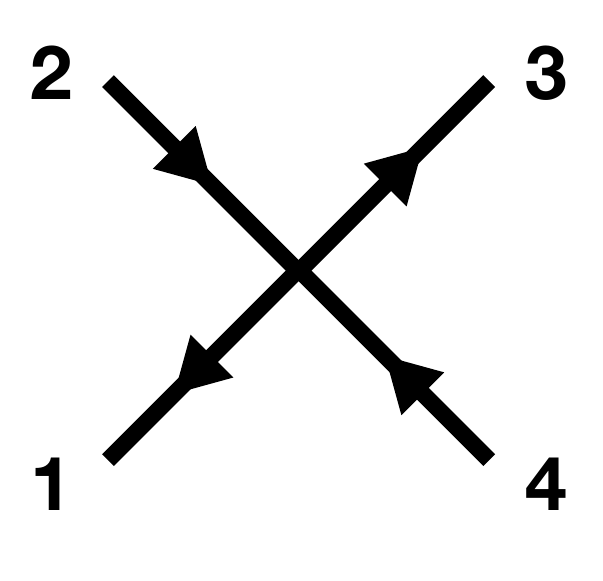}}} = \frac{1}{\Nf} \left[ \, F ( p_{12}^2 ) \, \delta_{12} \mkern1mu \delta_{34} - F ( p_{14}^2 ) \, \delta_{14} \mkern1mu \delta_{32} \, \right] \,.
\ee
We use the shorthand notation $\delta_{12} \equiv \delta_{a_1 a_2} \delta_{\alpha_1 \alpha_2}$ where numerical labels refer collectively to flavour indices, spinor indices, and external momenta directed to follow the fermion arrows. The vertex also depends on combinations of momenta $p_{ij} = p_i - p_j$, with $p_{12}$ and $p_{14}$ appearing separately in the two distinct tensor structures.

At one-loop, diagrams show three basic topologies
\be\label{eq:splitLoops}
\vcenter{\hbox{\includegraphics[width=.85\linewidth]{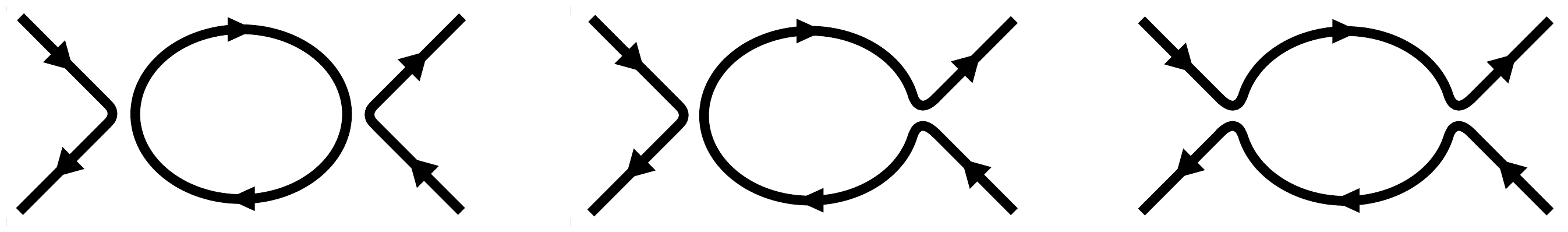}}}
\ee
where split vertices reflect   the distinct index contractions in \eqref{eq:4Frule}. Only the first type of  contraction leads to a flavour trace $\propto \Nf$ and gives the large-$\Nf$ leading contribution. Since Dirac indices follow the same contractions, it also follows that the  tensor structures from  leading diagrams are products of Kronecker deltas such as those in the tree-level vertex. 

Moreover, each vertex  in \eqref{eq:splitLoops} now carries a factor $F ( p^2 )$, where $p$ is equal to the sum of (incoming) momenta on either pair of continuously connected lines. Hence, vertices  in diagrams of the first type only depend on external momenta, while those in diagrams of the second and third type also depend  on loop momenta. It follows that  all large-$\Nf$ leading diagrams factorise external momenta from loop momenta  in the form $F(p^2)^2 \times B(p^2)$. Applying this observation to the entire bubble sum  \eqref{eq:bubbleSum} at leading order in $1/\Nf$, we conclude that the uplift of the vertex neatly feeds through, leading to the replacement of $G$ in the geometric series by the function $F$.

With these insights at hand, we sum over all bubbles, including crossed diagrams, to find the leading-order four-point function as 
\be\label{eq:4Fdressed}\nonumber
\vcenter{\hbox{\includegraphics[height=4em]{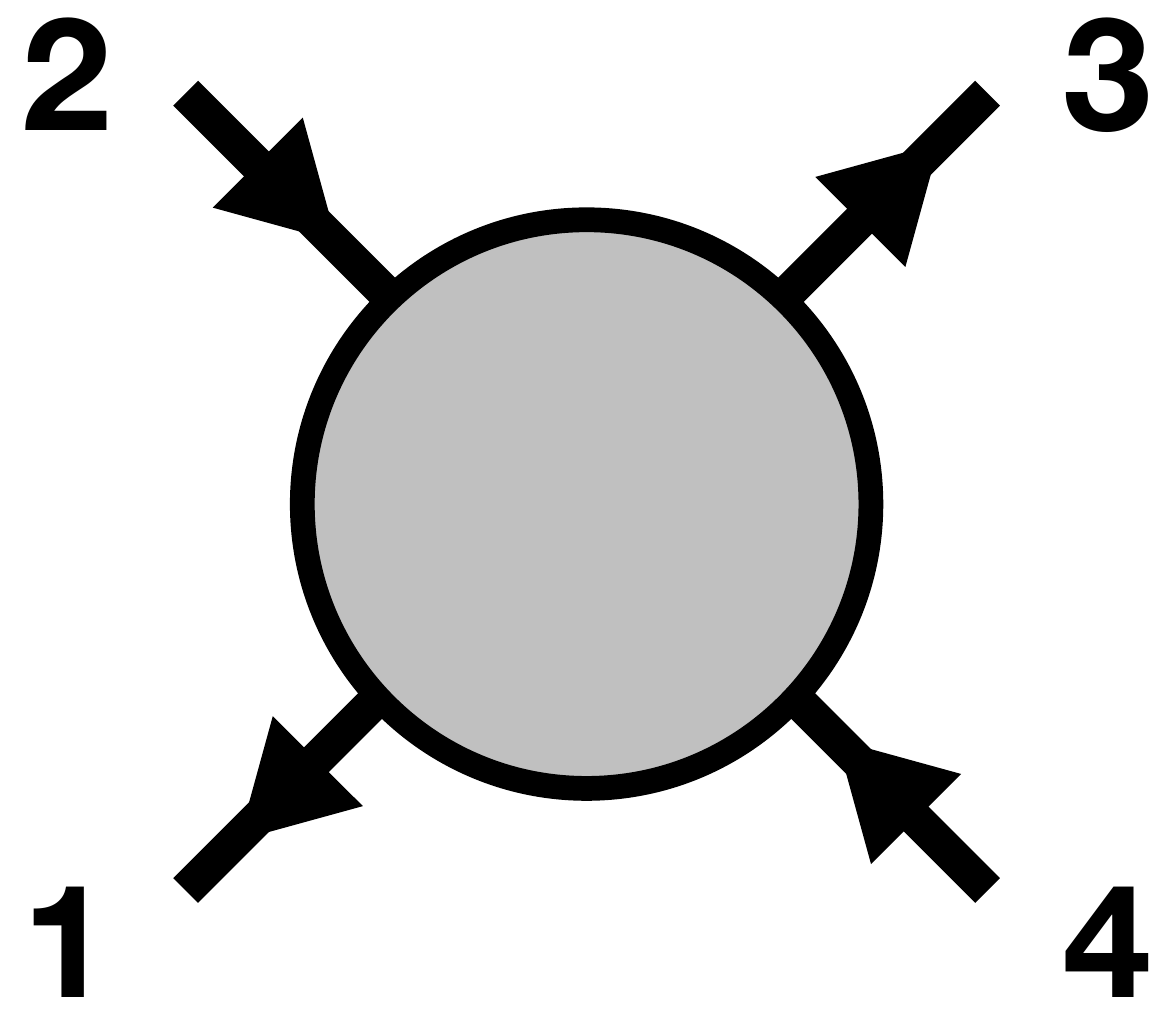}}} = \frac{1}{\Nf} \left[ \, V ( p_{12}^2 ) \, \delta_{12} \mkern1mu \delta_{34} - V ( p_{14}^2 ) \, \delta_{14} \mkern1mu \delta_{32} \, \right] \,.
\ee 
The result looks identical to the tree-level vertex only with $F$ uplifted into the resummed vertex function $V$,
\be\label{eq:V4d}
V ( p^2 ) = \frac{1}{Z p^2 + G^{-1} + B ( p^2 )} \,.
\ee
Crucially, the result \eqref{eq:V4d}  makes it evident that the four-point vertex can be rendered finite by absorbing the two types of UV divergences from the bubble integral \eqref{eq:Bdiv4d}  into the parameters $Z$ and $G^{-1}$. We also conclude that the counterterms for infinitely many higher-derivative interactions are fixed in terms of just two parameters.

\begin{figure}[!t]
\includegraphics[width=.8\linewidth]{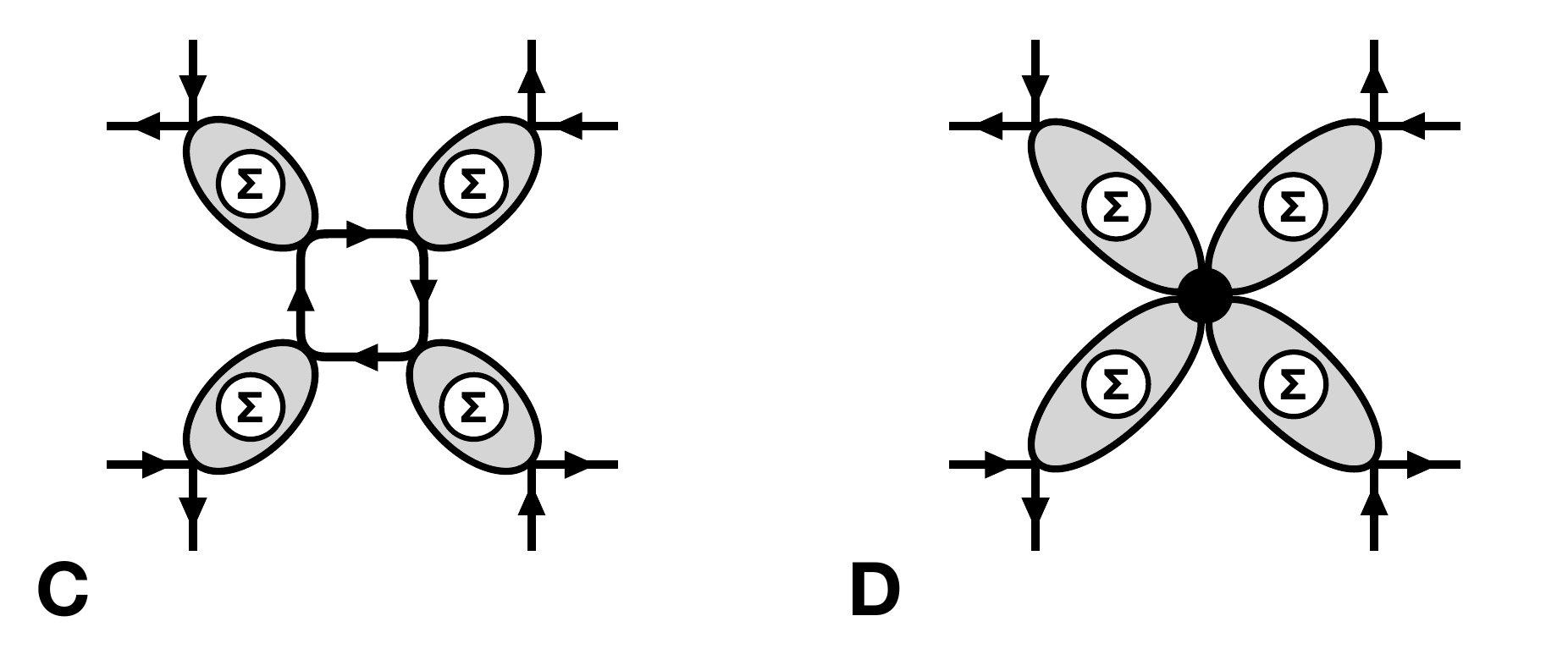}
\caption{Large-$\Nf$ leading contributions to the 8F vertex in the extended GN theory. Shaded blobs with  label $\Sigma$ represent an infinite series of large-$\Nf$ fermion bubbles given in \eqref{eq:4Fsum}.}
\label{fig:8Fdiags}
\end{figure}

{\it Renormalising the 8F vertex.---} 
We proceed with finding an elementary 8F vertex  able to absorb the logarithmic divergence \eqref{eq:Ddiv4d}.
Its  form  can be appreciated graphically from the diagrams in Fig.~\ref{fig:8Fdiags}, representing the large-$\Nf$ leading contributions to the 8F vertex. Shaded blobs labelled $\Sigma$ in this figure denote the insertion of an infinite series of fermion bubbles,
\be\label{eq:4Fsum}
\vcenter{\hbox{\includegraphics[height=2.5em]{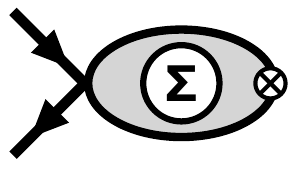}}} 
= \vcenter{\hbox{\includegraphics[height=2.5em]{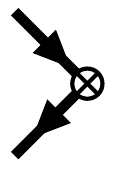}}} 
+ \vcenter{\hbox{\includegraphics[height=2.5em]{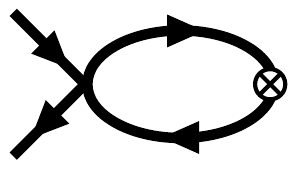}}} 
+ \cdots
\ee
where, following the discussion after \eqref{eq:splitLoops}, only large-$\Nf$ leading index contractions are implied. For the box subgraph in diagram C, this means that only contractions with a trace over all four propagators in the loop are taken into account. 

First consider diagram C in Fig.~\ref{fig:8Fdiags}, containing the box subgraph. Each bubble chain attached to the box carries a factor $1/\Nf$, while the central fermion loop gives a factor $\Nf$, meaning that the leading-order power counting for 8F diagrams is $1/\Nf^3$. Just as for the 4F vertex, the momentum dependence of the vertices factorises from the loop momenta in the large-$\Nf$ leading diagrams.  
Summing up the bubble chains results in terms of the form
\be\label{eq:diaC}
-\tfrac{1}{N^3} V ( p_1^2 ) \cdots V ( p_4^2 ) \, D ( p_1, p_2, p_3 ) \, ,
\ee
where $p_i$ are the momenta carried by each bubble chain, subject to momentum conservation $\sum_i p_i = 0$, and $D(p_i,p_j,p_k)$ is the fermion box integral. Each of these  come with a tensor structure $\delta_{12} \delta_{34} \delta_{56} \delta_{78}$ of which there are $4!$~signed index permutations, times $3!$~un\-signed permutations of the  momenta in $D(p_1,p_2,p_3)$. Subdivergences in  $V$ are taken care of by the renormalisation of $Z$ and $G^{-1}$. With \eqref{eq:Ddiv4d}, the remaining overall divergent structure is of the form $\propto \tfrac{1}{N^3} V ( p_1^2 ) \cdots V ( p_4^2 ) \ln \Lambda$. 

Now consider diagram D in Fig.~\ref{fig:8Fdiags}. The central black dot stands for the 8F vertex 
that needs to absorb the  divergence from diagram C. Following the discussion around \eqref{eq:diaC} and modulo permutations, 
the vertex takes the form $F_8 ( p_1, p_2, p_3 ) \, \delta_{12} \delta_{34} \delta_{56} \delta_{78}$. It then remains to determine the function $F_8$. Using \eqref{eq:4Fsum}, we find that diagram D involves permutations of terms like
\be\nonumber
F_8 ( p_1, p_2, p_3 ) \times \prod_{i=1}^4 \sum_{n=0}^\infty [-F(p_i^2) B(p_i^2)]^n \, .
\ee
Each bubble sum yields $V(p_i^2) F ( p_i^2 )^{-1}$, and a factor $\tfrac{1}{N} F ( p_i^2 )$ has been removed from each compared to \eqref{eq:diaC}. 
To cancel the divergence in diagram C, the function $F_8$ must restore all four factors of $F ( p_i^2 )$ and three factors of $\tfrac{1}{\Nf}$. All in all, the quasi-local 8F vertex in position space reads 
\be\label{eq:8ptResum}
\frac{\lambda}{4! \Nf^3} \left[ F ( -\partial^2 ) (\psib_a\psi_a) \right]^4 \, 
\ee
with $\lambda$ a dimensionless 8F coupling of order unity  that absorbs the logarithmic divergence \eqref{eq:Ddiv4d}. 

The  elementary vertex \eqref{eq:8ptResum} consists of a primary 8F operator with pointlike $(\psib\psi)^4$ interactions, and  infinitely many higher-derivative 8F Mandelstam descendants of the form $\sim \partial^{2i}(\bar\psi\psi) \partial^{2j}(\bar\psi\psi) \partial^{2k}(\bar\psi\psi) \partial^{2\ell}(\bar\psi\psi)$ inherited from the 4F vertex. The dimensionful coupling associated to the primary operator is $\bar\lambda_{8 \rm F} = G^4 \lambda$. Given that $[\lambda]=d-4$, its mass dimension $[\bar\lambda_{8 \rm F}]$ originates entirely from the primary 4F interaction $[G]$.

{\it No divergences from higher-point vertices.---} 
Since each loop can have at most one flavour trace, large-$\Nf$ leading diagrams are exactly those where the number of flavour traces $T$  equals the number of momentum loops $L$. Hence, leading diagrams consist of products of non-overlapping one-loop subdiagrams, no matter the number of loops they contain. This must be so since flavour traces correspond to closed index loops that cannot share lines, while momentum loops can, e.g.~\eqref{eq:splitLoops}. Overlapping subgraphs, by definition, are those which share at least one internal line, and  any diagram containing these  has $T < L$ and is subleading. Given that divergences of tadpole and triangle diagrams vanish by chiral symmetry, and that diagrams with more than four internal  propagators are convergent, we conclude that the subdiagrams in Fig.~\ref{fig:subdivs} are the only sources of divergences. Together with 4F vertices always appearing as fully resummed bubble chains and 8F vertices appearing fully dressed with bubble chains and dressed box diagrams (Fig.~\ref{fig:8Fdiags}), it is guaranteed that the renormalisation of the three parameters $Z$, $G$ and $\lambda$  suffices to render all diagrams finite.

{\it Extended Gross-Neveu theory.---}
Putting all pieces together, we conclude that a renormalisable extension of \eqref{eq:localGN} with elementary 4F and 8F interactions is given by 
\be\label{eq:nonlocalGN}
\begin{split}
L_{\rm eGN}=Z_\psi\psib_a \slashed{\partial} \psi_a - \tfrac{1}{2 N} ( \psib_a \psi_a ) F ( -\partial^2 ) ( \psib_b \psi_b ) \\ + \tfrac{\lambda}{4! N^3} [ F ( -\partial^2 ) ( \psib_a \psi_a ) ]^4 \, .
\end{split}
\ee
The theory is fundamentally characterised by the 4F coupling $G$, dimensionless higher-derivative and 8F couplings $Z$ and $\lambda$, and a wave-function factor $Z_\psi\neq 1$ starting at order  $1/\Nf$ due to the sunset diagram. The free theory is recovered for $G\to 0$. The reasoning developed previously in three dimensions  \cite{Rosenstein:1988pt,Rosenstein:1990nm} can now be used in four dimensions to establish renormalisability to all orders in the $1/\Nf$ expansion \cite{future}.

\begin{figure*}
\vskip-.4cm
\includegraphics[width=1\linewidth]{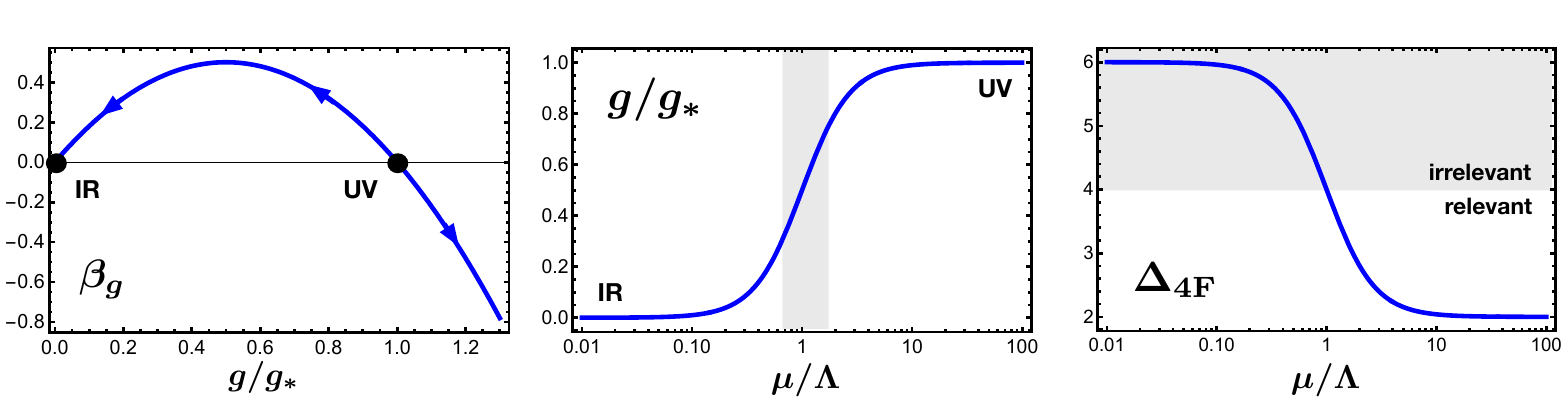}
\vskip-.5cm
\caption{Shown are the beta function \eqref{eq:betag} and its free and interacting fixed points (left), and the 4F coupling  \eqref{eq:g}  (middle)  and scaling dimension $\Delta_{\rm 4F}$ \eqref{eq:Delta}  (right) as functions of the renormalisation group  scale $\mu$  with $\Lambda=(G_F/g_*)^{-1/2}$.}\label{fig:fig3}
\vskip-.4cm
\end{figure*}

{\it Universality and the renormalisation group.---} 
It is useful to  understand the renormalisability of \eqref{eq:nonlocalGN} from the viewpoint of the renormalisation group.
Setting renormalisation conditions at the scale $\mu$, physical observables solely depend on  dimensionless running couplings  $Z(\mu)$, $\lambda(\mu)$ and $g ( \mu ) \equiv \mu^2G ( \mu )$. Their large-$\Nf$ exact beta functions are found using renormalisation schemes that are sensitive to  the underlying divergences   \eqref{eq:Bdiv4d}, \eqref{eq:Ddiv4d}  \cite{Cresswell-Hogg:2025wda,PDS}. The running of  $Z$ and $\lambda$ relate to logarithmic divergences and give universal  beta functions 
\be\label{eq:betaZla}
\beta_Z 
= -\frac{1}{4\pi^2} \, ,\quad \ 
\beta_\lambda 
= -\frac{3}{\pi^2} \,.
\ee
Both couplings display  a slow logarithmic running with  fixed points  $Z^{-1}=0=\lambda^{-1}$  that are infrared  or asymptotically free, depending on the signs of $Z$ and $\lambda$. The running of $g$ originates from  quadratic divergences \eqref{eq:Bdiv4d}, 
\be\label{eq:betag}
\mu\frac{dg}{d\mu}\equiv\beta_g = 2 g \left( 1 - \frac{g}{g_*} \right) \,,
\ee
implying that $g_*$ is genuinely non-universal, taking different values in different schemes.\footnote{For instance $g_* = 8 \pi^2$ in a momentum subtraction scheme.} We observe a Gaussian  fixed point $(g=0)$ at low energies and an interacting  fixed point $(g^{-1}_*\neq 0)$ at high energies, see Fig.~\ref{fig:fig3}, and
\be\label{eq:g}
g(\mu)=\frac{G_F\,\mu^2}{1+G_F\,\mu^2/g_*}
\ee
with $G_F\equiv G(0)$ interpolating inbetween. It follows that the Fermi scale $G_F^{-1/2}$ characterises the crossover from strong to weak coupling.

The interacting UV fixed point   clarifies why the theory has become  renormalisable in Wilson's sense. 4F interactions  in four space-time dimensions have a large scale dimension $\Delta_{\rm 4F}=4+\tfrac{\partial\beta_g}{\partial g}|_{g=0}=6$  classically. 
In the interacting theory, however, $\Delta_{\rm 4F}$ receives    quantum  corrections (Fig.~\ref{fig:fig3}, right panel) that  ultimately turn irrelevant  4F interactions into relevant ones,
\be\label{eq:Delta}
\Delta_{\rm 4F}|_*= 4+\left.\frac{\partial\beta_g}{\partial g}\right|_{g=g_*}= 2+{\cal O}\left(\frac1{\Nf}\right) \,.
\ee
The result \eqref{eq:Delta} is universal, and independent of the value $g_*$. Similarly, and in line with the structure of  the vertex \eqref{eq:8ptResum}, the classically irrelevant 8F interactions are rendered marginal $\Delta_{\rm 8F}|_*=4+{\cal O}(\tfrac1{\Nf}) $ by  quantum effects inherited from the renormalised 4F vertex. 

We conclude that the UV theory  exhibits  a  conformal critical point $(g=g_*)$ with a relevant primary 4F operator, and marginal descendant 4F and 8F operators responsible for  a mild violation of quantum scale symmetry at finite $Z$ and $\lambda$. Overall, the theory \eqref{eq:nonlocalGN}  is as predictive and well-behaved as perturbatively renormalisable ones, e.g.~QED, Yukawa theories, or  QCD.

{\it Discussion.---} 
We have demonstrated that  canonically non-renormalisable theories in four space-time dimensions may very well be renormalisable  in Wilson's sense. The reason for this are  quantum effects that turn  canonically irrelevant interactions into relevant or marginal ones, much in the spirit of an asymptotic safety conjecture \cite{Weinberg:1980gg}.  Also, higher-derivative interactions are mandatory for the mechanism to be  operative in full. Ultimately, physical observables only depend  on a finite number of parameters, and Fermi's scale signals the crossover into a  strongly-interacting  regime  with radically modified operator scaling, Fig.~\ref{fig:fig3}, rather than a loss of predictivity. It will be  interesting to investigate the UV theory more extensively   \cite{future}  using,  for instance, functional renormalisation  \cite{Cresswell-Hogg:2025wda},  bosonisation   \cite{Rosenstein:1990nm,ZinnJustin:1991yn,Weinberg:1997rv}, perturbation theory  \cite{PDS}, field redefinitions  \cite{Pisarski:1983gn},   conformal field theory \cite{Hogervorst:2016itc},  the  bootstrap \cite{Karateev:2019pvw}, or  $4d$ lattice simulations.

It is also  interesting to take a ``bottom-up" perspective with  Fermi's scale  acting as  a UV cutoff. The standard procedure to renormalise \eqref{eq:localGN} in an effective theory expansion leads to  higher-dimensional  counterterms, ultimately infinitely many of them. Predictivity is recovered if   Wilson coefficients are matched to an underlying UV theory. In the light of \eqref{eq:nonlocalGN}, the perhaps unexpected  conclusion is that the renormalised  effective theory with suitably-tuned coefficients becomes its own UV completion, valid   above the Fermi scale  and with only a few free parameters. It will be illuminating to perform  a matching calculation in the  effective theory  to connect  ``bottom-up" with ``top-down", and to precisely understand the reduction of parameters  \cite{Zimmermann:1984sx,Kubo:2001tr}. 

In particle physics, the  landscape  of theories with  controlled interacting UV fixed points \cite{Litim:2014uca,Bond:2016dvk,Bond:2017suy,Bond:2017tbw,Bond:2018oco,Bond:2019npq,Bond:2021tgu,Hiller:2022hgt,Bond:2022xvr,Litim:2023tym,Steudtner:2024pmd,Bond:2025aht}  has been studied for some time \cite{Bond:2017wut,Hiller:2019mou,Hiller:2020fbu,Rischke:2015mea,Abel:2017ujy,Barducci:2018ysr,Bissmann:2020lge,Held:2020kze,Bajc:2020gpa,Bause:2021prv,Hiller:2022rla,Hiller:2024zjp}. What's new now is the discovery that the landscape also includes canonically irrelevant interactions that  become UV-relevant due to quantum effects. This opens up new territory for  model building  beyond the confines of  perturbative renormalisability  \cite{Coleman:1973sx,Bond:2018oco}. For instance, with minor adjustments, our  proof extends to other types of fermionic  theories with  solvable large-$\Nf$ expansions such as~Thirring and Nambu--Jona-Lasinio models \cite{Thirring:1958in,Nambu:1961tp,Nambu:1961fr}. Further interesting directions relate to four-fermion interactions in  standard model extensions \cite{Grzadkowski:2010es,Ferretti:2013kya}, in strongly-coupled gauge theories \cite{Aoki:1996fh,Aoki:1999dv, Kubota:1999jf, Hasenfratz:2022qan},  in models without a fundamental Higgs  \cite{Bardeen:1989ds,Hasenfratz:1991it,Gies:2003dp}, with the symmetries of QCD \cite{Braun:2011pp,Gies:2005as}, or   in the context of quantum gravity \cite{Eichhorn:2011pc}.  In light of our results, their designation as effective rather than fundamental could  be re-evaluated. We  look forward to more extensive studies of  \eqref{eq:nonlocalGN}  and its cousins \cite{future}.

{\it Acknowledgements.---} We thank the participants of the 12th workshop {\it Asymptotic Safety meets Particle Physics and Friends} (DESY, Hamburg, Dec 2025) for discussions. This work is supported by the Science and Technology Facilities Council (STFC)  under the Consolidated Grant ST/T00102X/1.

\bibliography{4dLargeN}
\bibliographystyle{mystyle}

\end{document}